\documentclass[aps,prc,onecolumn,showpacs,preprintnumbers,amsmath,amssymb]{revtex4}

\usepackage{dcolumn}
\usepackage{bm}

\usepackage{amsmath,amssymb,graphicx}

\usepackage{multirow}
\usepackage{amsfonts}
\usepackage{dsfont}
\usepackage{subfigure}
\usepackage{epsfig}
\usepackage{graphicx}
\usepackage{graphics}
\usepackage{subfigure}
\usepackage{verbatim}

\usepackage{color}
\begin{document}
\title{A Study of P-wave Heavy Meson Interactions in A Chiral
Quark Model}
\author{M.T. Li$^{1,2}$}
\author{W.L. Wang$^{3}$}
\author{Y.B. Dong$^{1,2}$}
\author{Z.Y. Zhang$^{1,2}$}
\affiliation
{\small $^1$Institute of High Energy Physics, P.O. Box 918-4, Beijing 100049, China\\
$^2$Theoretical Physics Center for Science Facilities (TPCSF),
CAS, Beijing 100049, China \\
$^3$Physics and Nuclear Energy Engineering, Beihang University,
Beijing 100191, China}

\date{\today}

\begin{abstract}
The analytical forms of the interaction potentials between one
S-wave and one P-wave heavy mesons as well as the potentials between
two P-wave heavy mesons are deduced based on a chiral quark model.
Our results explicitly show the attractive property between two
heavy mesons. Consequently, a series of possible molecular states
are obtained. It is expected that our study might shed some light on
the popular discussions of the newly observed XYZ states.
\end{abstract}


\maketitle

\section{introduction}\label{sec1}
The newly discovered XYZ states, such as new charmonium-like states
of $X(3872)$ \cite{3872-Belle,3872-CDF,3872-D0,3872-Babar},
$X(3940)$ \cite{x3940-Belle,y3940-Belle,y3940-Babar},
$X(4050)^{\pm}$ \cite{4050-Belle,4050-Babar}, $X(4140)$
\cite{4140-Belle,4140-CDF}, $X(4160)$
\cite{X4160-Belle},
$X(4250)^{\pm}$ \cite{4050-Belle,4050-Babar}, $X(4260)$
\cite{4260-Babar,4260-Cleo,4260-Belle}, $X(4350)$ \cite{4140-Belle},
$X(4360)$ \cite{4360-Belle,4360-Babar}, $X(4430)^{\pm}$
\cite{4430-Belle,4430-Belle-again,4430-Babar}, $X(4660)$
\cite{4360-Belle,4360-Babar}, and  botomnium-like states of
$Z_b(10610)^{\pm}$ \cite{Zb-Belle}, $Z_b(10650)^{\pm}$
\cite{Zb-Belle}, are of great interests, since they cannot be simply
considered as normal quarkonium of $c\bar{c}$ or $b\bar{b}$,
especially for the charged ones. For these XYZ states, many
different explanations have been taken into account, including
tetraquark, molecular states, hybrid charmonia, cusps, and threshold
effects.

In our previous works \cite{li1,li2}, the possible two S-wave heavy
meson molecular explanations for the new resonances of $X(3872)$,
$X(3940)$, $Z_b(10610)^{\pm}$ and $Z_b(10650)^{\pm}$ are obtained.
In this paper we extend our study of the S-wave heavy meson
interactions to the P-wave heavy meson interactions, and we try to
explain the structures of the other XYZ states, particularly for
$X(4250)^{\pm}$, $X(4350)$, $X(4360)$, $X(4260)$,
$X(4430)^{\pm}$. As far as $X(4250)^{\pm}$ is concerned, it was
explained as a $I^G(J^{PC})=0^+(1^{-})$ $D^*\bar{D}_1$ molecule by
Close \cite{close}, however Nielsen et al.
\cite{nielsen,nielsen1,nielsen2} considered it as a $1^+(1^{-})$
$D\bar{D}_1$ molecule with QCD sum rules, while Ding \cite{ding}
found that the $D\bar{D}_1$ and $D^*_0\bar{D}^*$ molecules couldn't
be the explanation for $X(4250)^{\pm}$, but for $X(4260)$. For the
latter resonance, Close \cite{close} and Nielsen et al.
\cite{nielsen, nielsen1,nielsen2} both agreed with Ding's
conclusion. Moreover, Nielsen et al.
\cite{nielsen,nielsen1,nielsen2} concluded that $X(4430)^{\pm}$
could be regarded as a tetraquark or a $D^*\bar{D}_1$ molecule.
Close \cite{close} and Ding et al. \cite{ding1} confirmed this
molecular explanation, however, the calculation of Liu et al.
\cite{liu0} disfavored the $D^*\bar{D}_1$ molecular explanation
using the resonating group method in our chiral quark model.

In this paper, our chiral quark model will be employed to study the
heavy meson interactions. The model has been explained, in detail,
in Refs.\cite{zhang1, zhang2, dai, fhuang04kn, fhuang04nkdk,
liu1,liu2,wangwlsigc}. In this chiral SU(3) quark model,
one-gluon-exchange (OGE), confinement potential, scalar meson
exchange as well as pseudoscalar meson exchange are taken into
account. Moreover, this chiral SU(3) quark model was developed into
the extended chiral SU(3) quark model by introducing the vector
meson exchanges. It has been proved that both models are quite
successful in reproducing the spectra of the baryon ground states,
the binding energy of deuteron, the nucleon-nucleon ($NN$),
kaon-nucleon ($KN$) scattering phase shifts, and the hyperon-nucleon
($YN$) cross sections \cite{zhang1, zhang2, dai, fhuang04kn,
fhuang04nkdk}. During the past years, by using the Resonating Group
Method (RGM), Liu \cite{liu0,liu1,liu2} and Wang \cite{wangwlsigc}
have studied the interactions and structures of the heavy-quark
systems in both the chiral SU(3) quark model and the extended one.

Different from the RGM method, we have deduced analytical effective
interaction potentials between the two S-wave mesons in our chiral
quark models \cite{li,li1,li2}. In this paper, we try to go further
to deduce the analytical effective interaction potentials between
one S-wave and one P-wave heavy mesons as well as between two P-wave
heavy mesons. Then, we'll systematically calculate the possible
bound states associated with $D_0^*$, $D_1$, $D_2^*$ and their
bottom partners $B_1$, $B_2^*$, by solving the Schr\"{o}dinger
equation with our analytical potentials, and try to explain the
structures for some newly observed XYZ states, particularly for
$\psi(4160)$, $X(4250)^{\pm}$, $X(4350)$, $X(4356)$, $X(4260)$, and
$X(4430)^{\pm}$. We expect our approach could give a more accurate
description of the short-range heavy meson-meson interactions than
the RGM does, especially comparing with Ref. \cite{liu0}.

The paper is organized as follows. In section \ref{sec:formulism},
the framework of our chiral quark models is briefly introduced, and
the analytical forms of the effective interaction potentials between
one S-wave meson and one P-wave meson (S-P) and between two P-wave heavy
mesons (P-P) in our chiral quark models are given. The bound state
solutions are shown and discussed in Sec.~\ref{sec:result}. Finally,
a short summary is given in Sec.~\ref{sec:sum}.

\section{The chiral quark model}\label{sec:formulism}

The framework of our models has been discussed extensively in the
literature \cite{zhang1, zhang2, dai, fhuang04kn, fhuang04nkdk,
liu1,liu2,wangwlsigc, wang2007,wang2008,wang2010,wang2011}. As
mentioned in our previous papers \cite{li,li1,li2},  the internal
kinetic energies and the internal interactions of each meson are not
necessary to be calculated. The Hamiltonian of the relative motion,
in our approach, takes the form of
\begin{eqnarray}
H=T_{rel}+V_{eff},
\end{eqnarray}
where $T_{rel}$ is the kinetic energy operator of the relative
motion between the two mesons, and $V_{eff}$ is the effective
interaction potential derived from the quark-quark (quark-antiquark)
interaction between two mesons by integrating out the internal
coordinates  $\vec{\xi}_1$ and $\vec{\xi}_2$ of the two mesons:
\begin{eqnarray}\label{veff}
V_{eff}=\sum_{ij}\int\varphi_1^*(\vec{\xi}_1)
\varphi_2^*(\vec{\xi}_2)V(\vec{r}_{ij})\varphi_1(\vec{\xi}_1)
\varphi_2(\vec{\xi}_2) d\vec{\xi}_1d\vec{\xi}_2,
\end{eqnarray}
while $\varphi_1(\vec{\xi}_1)$ and $\varphi_2(\vec{\xi}_2)$ are the
intrinsic wavefunctions of the two mesons. If a meson is in S-wave,
its wave function is taken as one-Gaussian form:
\begin{eqnarray}
\varphi(\vec{\xi})=\big(\frac{\mu\omega}{\pi}\big)^{3/4}
e^{-\frac{\mu\omega}{2}\xi^2},
\end{eqnarray}
and if in P-wave, it is
\begin{eqnarray}
\varphi_{1m}(\vec{\xi})=\dfrac{2\sqrt{2}}{3\pi^{\frac{1}{4}}}
(\mu\omega)^{5/4}\xi
e^{-\frac{\mu\omega}{2}\xi^2}Y_{1m}(\theta,\varphi).
\end{eqnarray}
Here, $\mu$ is the reduced mass of the two quarks inside each meson
cluster and $\omega$ is the harmonic-oscillator frequency of the
meson intrinsic wavefunction. $V(\vec{r}_{ij})$ in eq. (\ref{veff})
represents the interactions between the $i$-th light quark or
antiquark in the first meson and the $j$-th light quark or antiquark
in another.

Because there is no color-interrelated interaction between the two
color-singlet clusters, OGE interaction and confinement potential
between two mesons don't exist, we only consider the meson-exchange
interactions between the two meson clusters. In our chiral SU(3)
quark model, we have
\begin{eqnarray}
V(\vec{r}_{ij})=\sum^8_{a=0} V^{\sigma_a}(\vec{r}_{ij})+\sum^8_{a=0}
V^{\pi_a}(\vec{r}_{ij}),
\end{eqnarray}
and in the extended chiral SU(3) quark model
\begin{eqnarray}
V(\vec{r}_{ij})=\sum^8_{a=0} V^{\sigma_a}(\vec{r}_{ij})+\sum^8_{a=0}
V^{\pi_a}(\vec{r}_{ij})+\sum^8_{a=0} V^{\rho_a}(\vec{r}_{ij}),
\end{eqnarray}
$V^{\sigma_a}(\vec{r}_{ij})$ and $V^{\pi_a}(\vec{r}_{ij})$ are the
scalar meson exchange and pseudoscalar meson exchange interactions,
respectively. $V^{\rho_a}(\vec{r}_{ij})$ indicates the vector meson
exchange interactions. For quark-quark (antiquark-antiquark)
interaction, the explicit forms of $V^{\sigma_a}(\vec{r}_{ij})$,
$V^{\pi_a}(\vec{r}_{ij})$ and $V^{\rho_a}(\vec{r}_{ij})$ have been
described, in detail, in Refs. \cite{zhang1,zhang2,dai, fhuang04kn,
fhuang04nkdk,liu1,liu2,wangwlsigc,wang2007,wang2008,wang2010,
wang2011,li1} as
\begin{eqnarray}
V^{\sigma_a}(\vec{r}_{ij})&=& -C(g_{ch},m_{\sigma_a},\Lambda) X_1(m_{\sigma_a},\Lambda,r_{ij})
\left(\lambda^a_i\lambda^a_j\right),\label{scalar} \\
V^{\pi_a}(\vec{r}_{ij})&=& C(g_{ch},m_{\pi_a},\Lambda) \frac{m^2_{\pi_a}}{12m_im_j}
X_2(m_{\pi_a},\Lambda,r_{ij})  \left(\sigma_i\cdot\sigma_j\right)
\left(\lambda^a_i\lambda^a_j\right), \label{pseudoscalar}\\
V^{\rho_a}(\vec{r}_{ij}) &=& C(g_{\rm chv},m_{\rho_a},\Lambda)\Bigg[X_1(m_{\rho_a},
\Lambda,r_{ij}) + \frac{m^2_{\rho_a}}{6m_im_j} \left(1+\frac{f_{\rm chv}}
{g_{\rm chv}} \frac{m_i+m_j}{M_N}+\frac{f^2_{chv}}{g^2_{chv}}
\frac{m_im_j}{M^2_N}\right)   \nonumber \\
&& \times \,  X_2(m_{\rho_a},\Lambda,r_{ij}) \, (\sigma_i\cdot\sigma_j)\Bigg]
\left(\lambda^a_i\lambda^a_j\right),\label{vector}
\end{eqnarray}
with
\begin{eqnarray}
C(g_{ch},m,\Lambda) &=& \frac{g^2_{ch}}{4\pi} \frac{\Lambda^2}{\Lambda^2-m^2} m, \\
\label{ev1} X_1(m,\Lambda,r_{ij}) &=& Y(mr_{ij})-\frac{\Lambda}{m} Y(\Lambda r_{ij}), \\
\label{ev2} X_2(m,\Lambda,r_{ij}) &=& Y(mr_{ij})-\left(\frac{\Lambda}{m}\right)^3 Y(\Lambda r_{ij}), \\
Y(x) &=& \frac{1}{x}e^{-x},
\end{eqnarray}
where $\lambda^a$ is the Gell-Mann matrix in flavor space, and
$\Lambda$ is the cutoff mass indicating the chiral symmetry breaking
scale. In eqs. (7-13), $m_i$ and $m_j$  are the masses of the $i$-th
light quark or antiquark in the first meson and the $j$-th light
quark or antiquark in another, respectively,  while $m_{\sigma_a}$,
$m_{\pi_a}$ and $m_{\rho_a}$  in eqs. (\ref{scalar},
\ref{pseudoscalar}, \ref{vector}) are the masses of the scalar,
pseudoscalar and vector nonets, respectively. $M_N$ in eq.
(\ref{vector}) is a mass scale usually taken as the mass of nucleon
\cite{dai}. $g_{ch}$, in the above eqs., is the coupling constants
for the scalar and pseudoscalar nonets, while $g_{chv}$ and
$f_{chv}$ are the coupling constants for the vector coupling and
tensor coupling of vector nonets, respectively. For the
quark-antiquark interactions, the G-parity of the exchanged mesons
should also be taken into account.

In the two heavy meson systems, we don't consider the one-meson
exchange interactions between two heavy quarks or between one heavy
quark and one light quark, because these interactions are beyond our
SU(3) models. By using the method in Refs. \cite{li,li1,li2} and
integrating out the internal coordinates of two mesons as described
by eq. (\ref{veff}), we can get the total analytical effective
interaction potentials between two S-wave heavy mesons. Analogy to
the two S-wave systems, we can go further to deduce the effective
interaction potentials between one P-wave heavy meson and one S-wave
heavy meson and between two P-wave Heavy mesons. These effective
interaction potentials are listed as follows.
\begin{eqnarray}
V_{eff}(\vec{R})=\sum^8_{a=0} V_{q\bar{q}}^{\sigma_a}(\vec{R})+\sum^8_{a=0}
V_{q\bar{q}}^{\pi_a}(\vec{R})+\sum^8_{a=0} V_{q\bar{q}}^{\rho_a}(\vec{R}), \nonumber
\end{eqnarray}
with
\begin{eqnarray}
V_{q\bar{q}}^{\sigma_a}(\vec{R})&=& -G_{\sigma_a}C(g_{ch},m_{\sigma_a},\Lambda)
X_{1q\bar{q}}(m_{\sigma_a},\Lambda,R)\left(\lambda^a_q\lambda^a_{\bar{q}}
\right),\label{eff:scalar} \\
V_{q\bar{q}}^{\pi_a}(\vec{R})&=& G_{\pi_a}C(g_{ch},m_{\pi_a},\Lambda)
\frac{m^2_{\pi_a}}{12m_qm_{\bar{q}}} X_{2q\bar{q}}(m_{\pi_a},\Lambda,R)
 \left(\sigma_q\cdot\sigma_{\bar{q}}\right)\left(\lambda^a_q
\lambda^a_{\bar{q}}\right), \label{eff:pseudoscalar}\\
V_{q\bar{q}}^{\rho_a}(\vec{R}) &=& G_{\rho_a}C(g_{chv},m_{\rho_a},\Lambda)
\Bigg[X_{1q\bar{q}}(m_{\rho_a},\Lambda,R)
+ \frac{m^2_{\rho_a}}{6m_qm_{\bar{q}}}
\Big(1+\frac{f_{chv}}{g_{chv}}
\frac{m_q+m_{\bar{q}}}{M_N}+\frac{f^2_{chv}}{g^2_{chv}}
\frac{m_qm_{\bar{q}}}{M^2_N}\Big)   \nonumber \\
&& \times X_{2q\bar{q}}(m_{\rho_a},\Lambda,R) \, (\sigma_q\cdot
\sigma_{\bar{q}})\Bigg] \left(\lambda^a_q\lambda^a_{\bar{q}}\right).
\label{eff:vector}
\end{eqnarray}
Here, $m_q$ and $m_{\bar{q}}$ are masses of the light quark and
antiquark, respectively, $G_{\sigma_a,\pi_a,\rho_a}$ is the
$G$-parity of the exchanged meson, and $\vec{R}$ is the relative
coordinate between the two different mesons, namely, the relative
coordinate between the centers-of-mass coordinates of the two
mesons. Moreover in eqs. (14-16)
\begin{eqnarray}
\label{eff:ev1} X_{1q\bar{q}}(m,\Lambda,R) &=& Y_{q\bar{q}}(mR)
-\frac{\Lambda}{m} Y_{q\bar{q}}(\Lambda R), \\
\label{eff:ev2} X_{2q\bar{q}}(m,\Lambda,R) &=& Y_{q\bar{q}}(mR)
-\left(\frac{\Lambda}{m}\right)^3 Y_{q\bar{q}}(\Lambda R),
\end{eqnarray}
where the $Y_{q\bar{q}}(mR)$ is the modified Yukawa term.

For the two S-wave heavy meson interactions, as mentioned in
\cite{li,li1,li2}, the modified Yukawa term in eqs.
(\ref{eff:ev1},\ref{eff:ev2}) reads
\begin{eqnarray}
Y_{q\bar{q}}(mR)&=&\frac{1}{2mR}e^{\frac{m^2}{4\beta}}\bigg\{e^{-mR}
\Big\{1-erf\Big[-\sqrt{\beta}(R-\frac{m}{2\beta})\Big]\Big\}
-e^{mR}\Big\{1-erf\Big[\sqrt{\beta}(R+\frac{m}{2\beta})\Big]\Big\}\bigg\}.
\label{modified YukawaSS}
\end{eqnarray}
For the one S-wave heavy meson and one P-wave heavy meson
interactions, they receive the contributions from direct and
exchange terms. The modified Yukawa term of the direct term reads
\begin{eqnarray}
Y_{q\bar{q}}(mR)&=&\frac{1}{3mR}\Bigg\{ e^{\frac{m^2}{4\beta}}
\left(\frac{m^2_{\bar{Q}}m^2}{4\omega (m_{\bar{Q}}+m_q)^2\mu_{q\bar{Q}}}+\frac{3}{2}\right)
\bigg\{e^{-mR}
\Big\{1-erf\Big[-\sqrt{\beta}(R-\frac{m}{2\beta})\Big]\Big\}\nonumber \\
&&
-e^{mR}\Big\{1-erf\Big[\sqrt{\beta}(R+\frac{m}{2\beta})\Big]\Big\}\bigg\}-\beta^{\frac{3}{2}}R e^{-\beta
R^2}\frac{m^2_{\bar{Q}}}{\omega(m_{\bar{Q}}+m_q)^2\mu_{q\bar{Q}}}\Bigg\},
\label{modified YukawaSP}
\end{eqnarray}
and the one of the exchange term, associated with the charge-parity
$C$ in flavor wave functions (see eqs.
(\ref{wavef-dif1}-\ref{wavef-dif2}) in section III), reads
\begin{eqnarray}
Y_{q\bar{q}}(mR)&=&\frac{1}{3mR}\big(\mu_{q\bar{Q}}
\omega\big)^2\big(\mu_{\bar{q}Q}\omega\big)^2
\bigg(-\frac{m_{\bar{Q}}}{m_q+m_{\bar{Q}}}\frac{m_Q}{m_{\bar{q}}+m_Q}
\frac{\partial}{\partial a}+\frac{m_q+m_{\bar{Q}}}{4m_{\bar{Q}}}
\frac{m_{\bar{q}}+m_Q}{m_Q}\frac{\partial}{\partial b}\bigg)
\nonumber \\
&&\times\Bigg\{(a\beta)^{-\frac{3}{2}}e^{\frac{m^2}{4\beta}}\bigg\{e^{-mR}
\Big\{1-erf\Big[-\sqrt{\beta}(R-\frac{m}{2\beta})\Big]\Big\}
-e^{mR}\Big\{1-erf\Big[\sqrt{\beta}(R+\frac{m}{2\beta})\Big]\Big\}\bigg\}\Bigg\}.
\label{modified YukawaSPPS}
\end{eqnarray}
Moreover, for the two P-wave heavy meson interaction, the modified
Yukawa term is
\begin{eqnarray}
Y_{q\bar{q}}(mR)&=&\frac{2}{9mR}\Bigg\{ e^{\frac{m^2}{4\beta}}
\Big(\frac{m^2_{\bar{Q}}m^2}{4\omega (m_{\bar{Q}}+m_q)^2\mu_{q\bar{Q}}}
+\frac{3}{2}\Big)
\Big(\frac{m_Q^2m^2}{4\omega (m_Q+m_{\bar{q}})^2\mu_{\bar{q}Q}}+\frac{3}{2}\Big)
\nonumber \\
&&\times\bigg\{e^{-mR}
\Big\{1-erf\Big[-\sqrt{\beta}(R-\frac{m}{2\beta})\Big]\Big\}
-e^{mR}\Big\{1 -erf\Big[\sqrt{\beta}(R+\frac{m}{2\beta})\Big]\Big\}\bigg\}\nonumber \\
&&
+\frac{m^2_{\bar{Q}}}{(m_{\bar{Q}}+m_q)^2\mu_{q\bar{Q}}}
\frac{m^2_Q}{\omega^2(m_Q+m_{\bar{q}})^2\mu_{q\bar{Q}}}R e^{-\beta R^2}
\Big(-\frac{m^2}{4}\beta^{\frac{3}{2}}+\frac{3}{2}\beta^{\frac{5}{2}}+R^2\beta^{\frac{7}{2}}\Big)\Bigg\}\nonumber \\
&&-\frac{3\beta^{\frac{3}{2}}}{2\omega}R e^{-\beta R^2}
\Big[\frac{m^2_{\bar{Q}}}{(m_{\bar{Q}}+m_q)^2\mu_{q\bar{Q}}}+\frac{m^2_Q}{(m_Q+m_{\bar{q}})^2\mu_{q\bar{Q}}}\Big].\nonumber \\
\label{modified YukawaPP}
\end{eqnarray}
In the above equations,
\begin{eqnarray}
\beta&=&b-\frac{c^2}{4a},
\end{eqnarray} where
\begin{eqnarray}
a&=&\mu_{q\bar{Q}}\omega\left(\frac{m_Q}{m_{\bar{q}}+m_Q}\right)^2
+\mu_{\bar{q}Q}\omega\left(\frac{m_{\bar{Q}}}{m_q+m_{\bar{Q}}}\right)^2,\nonumber \\ \\
b&=&\mu_{q\bar{Q}}\omega\left(\frac{m_q+m_{\bar{Q}}}{m_{\bar{Q}}}\right)^2
+\mu_{\bar{q}Q}\omega\left(\frac{m_{\bar{q}}+m_Q}{m_Q}\right)^2,\nonumber \\ \\
c&=&\mu_{q\bar{Q}}\omega\frac{m_q+m_{\bar{Q}}}{m_{\bar{Q}}}\frac{m_Q}{m_{\bar{q}}+m_Q}
-\mu_{\bar{q}Q}\omega\frac{m_{\bar{q}}+m_Q}{m_Q}\frac{m_{\bar{Q}}}{m_q+m_{\bar{Q}}},
\end{eqnarray}
and $m_Q$ and $m_{\bar{Q}}$ the are masses of the heavy quark and
antiquark, respectively, and $\mu_{qQ}=\frac{m_qm_Q}{m_q+m_Q}$.

In this work, we take our model parameters, which have been
determined in our previous works, as follows
\cite{dai,liu1,liu2,zhang1,fhuang04kn,fhuang04nkdk,zhang2,wangwlsigc,
wang2007, wang2008, wang2010}. The up/down quark mass $m_q$ is
fitted as the nucleon mass and taken as $M_N/3\sim 313$ MeV. The
coupling constant for the scalar and pseudoscalar chiral fields
$g_{ch}=2.621$ is fixed by the relation of
\begin{eqnarray}
\frac{g^{2}_{ch}}{4\pi} =\frac{9}{25} \frac{g^{2}_{NN\pi}}{4\pi}
\frac{m^{2}_{u}}{M^{2}_{N}},\nonumber
\end{eqnarray}
with $g^{2}_{NN\pi}/4\pi=13.67$ determined from experiments.

In our extended chiral SU(3) quark model, the vector coupling
constant $g_{chv}$ and tensor coupling constant $f_{chv}$ in eqs.
(\ref{vector}, \ref{eff:vector}) are fitted by the mass difference
between $N$ and $\Delta$, under the condition that the strength of
the OGE is taken to be almost zero. When the tensor coupling is
neglected, $g_{chv}=2.351$ and $f_{chv}=0$; and when the tensor
coupling is considered, $g_{chv}=1.973$ and $f_{chv}=1.315$. The
harmonic-oscillator frequency $\omega$ (being as $1/(m_u
b_u^2)=1/(m_c b_c^2)$ where $b_u$ is fitted by the $N-N$ scattering
phase shifts) is taken as $2.522 fm^{-1}$ in the chiral SU(3) quark
model and $3.113 fm^{-1}$ in the extended chiral SU(3) quark model,
respectively. In our calculation, the masses of the mesons are taken
from the PDG \cite {PDG2010}, except for the $\sigma$ meson, which
does not have a well-defined value. Here $m_\sigma$ is obtained by
fitting the binding energy of deuteron \cite{dai}. It is
$m_\sigma=595$ MeV in our chiral SU(3) quark model, 535 MeV for
neglecting tensor coupling and 547  MeV for considering tensor
coupling in our extended chiral SU(3) quark model. The cutoff mass
$\Lambda$ means the chiral symmetry breaking scale and is taken as
$1100$ MeV as a convention. In addition, we find that the final
results are not sensitive to the variation of the heavy quark
masses, and we take  $m_c=1430$  MeV \cite{zhanghx1} and $m_b=4720$
MeV \cite{zhanghx2} as the two typical values.

\section{Results and discussions}\label{sec:result}

To make sure the two-heavy meson systems have definite quantum
numbers of isospin $I$ and $C$-parity $C$, we follow the definitions
of the flavor wavefunctions in Refs. \cite{liu1,liux,he,li1}.
For the hidden-charm systems constituted by two different mesons
$\mathfrak{D}^{'}$ and $\bar{\mathfrak{D}}$, the flavor wavefunctions are

\begin{eqnarray}
&I=1:&\begin{cases}
\dfrac{1}{\sqrt{2}}(\mathfrak{D}^{'+}\bar{\mathfrak{D}}^0+c\mathfrak{D}^{+}\bar{\mathfrak{D}}^{'0})\\
\dfrac{1}{\sqrt{2}}(\mathfrak{D}^{'-}\bar{\mathfrak{D}}^0+c\mathfrak{D}^{-}\bar{\mathfrak{D}}^{'0})\\
\dfrac{1}{2}[(\mathfrak{D}^{'0}\bar{\mathfrak{D}}^{0}-\mathfrak{D}^{'+}\mathfrak{D}^{-})
+c(\mathfrak{D}^{0}\bar{\mathfrak{D}}^{'0}-\mathfrak{D}^{+}\mathfrak{D}^{'-})],\label{wavef-dif1}
\end{cases} \\
&I=0:&\frac{1}{2}[(\mathfrak{D}^{'0}\bar{\mathfrak{D}}^{0}+\mathfrak{D}^{'+}\mathfrak{D}^{-})
+c(\mathfrak{D}^{0}\bar{\mathfrak{D}}^{'0}+\mathfrak{D}^{+}\mathfrak{D}^{'-})].\label{wavef-dif2}
\end{eqnarray}
Here, $c$=1 for $C$=+ and $c$=--1 for $C$=--. As for  $\mathfrak{D}$
meson-anti $\bar{\mathfrak{D}}$ meson systems, the flavor functions
are
\begin{eqnarray}
&I=1:&\begin{cases}
\mathfrak{D}^{+}\bar{\mathfrak{D}}^{0}\\
\mathfrak{D}^{-}\bar{\mathfrak{D}}^{0}\\
\dfrac{1}{\sqrt{2}}(\mathfrak{D}^{0}\bar{\mathfrak{D}}^{0}
-\mathfrak{D}^{+}\mathfrak{D}^{-}),
\end{cases} \\
&I=0:&\dfrac{1}{\sqrt{2}}(\mathfrak{D}^{0}\bar{\mathfrak{D}}^{0}+\mathfrak{D}^{+}\mathfrak{D}^{-}).
\end{eqnarray}
For the two $\mathfrak{B}$ systems, the similar expressions can be
obtained.

\subsection{Two D masons}

\subsubsection{S-P: $D\bar{D}_1$, ($D\bar{D}^*_0$, $D\bar{D}^*_2$, $D^*\bar{D}_1$) and
($D^*\bar{D}^*_0$, $D^*\bar{D}^*_2$)}

\begin{figure}[htb]
\centering
\includegraphics[scale=0.3]{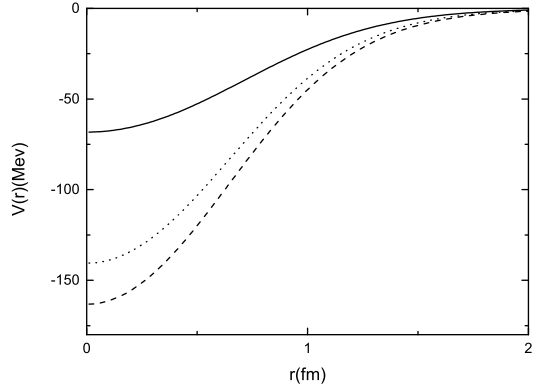}
\caption{The total interaction potential of $I$=0 and  $C$=--
$D\bar{D}_1$ system.  The solid, dashed and dotted lines represent
the results obtained from the chiral SU(3) quark model, the extended
chiral SU(3) quark model neglecting and including the tensor
coupling of vector field, respectively. \label{SP:SS} }
\end{figure}

For the two  pseudoscalar meson system of $D\bar{D}_1$, the spin
factor $<\sigma_q\cdot\sigma_{\bar{q}}>$ in eqs.
(\ref{eff:pseudoscalar}, \ref{eff:vector}) is 0. Therefore, the
spin-correlated interactions don't exist. Its interaction is
associated with isospin $I$ and $C$-parity $C$.
We find two bound states, and their binding energies are shown in
Table \ref{SPSV}.  In the case of $I=0$ $C=+$, the $I^G(J^{PC})=0^+(1^{-+})$
$D\bar{D}_1$  bound state  has a mass of 4253-4285 MeV.
In the case of $I=0$ $C=-$, the total interaction potential is shown in Fig. \ref{SP:SS},
 and the $0^-(1^{--})$
$D\bar{D}_1$  bound state has a mass of 4264-4285 MeV,
Its mass and quantum umbers match the $X(4260)$.
So in our approach, the $0^-(1^{--})$ $D\bar{D}_1$ molecule might be an
explanation to the state of $X(4260)$.

\begin{figure}[htb]
\centering
\includegraphics[scale=0.3]{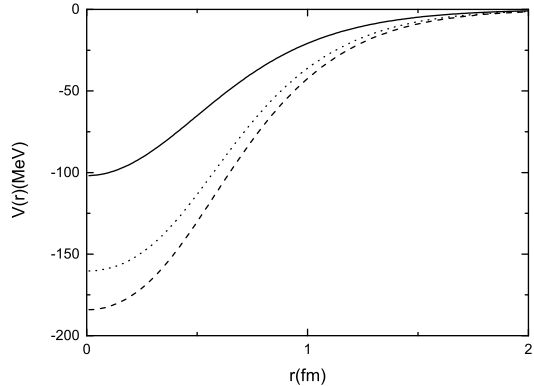}
\caption{The total interaction potential of $I$=0 and  $C$=+
$D\bar{D}^*_0$, $D\bar{D}^*_2$ and $D^*\bar{D}_1$ systems.  The solid, dashed
and dotted lines represent the same meaning as in Fig. \ref{SP:SS}.\label{SP:SV} }
\end{figure}

For the systems of  $D\bar{D}^*_0$, $D\bar{D}^*_2$ and
$D^*\bar{D}_1$, because they share the same interaction potential
with the same isospin and $C$-parity, we put them together in
discussion. We find the total interaction potentials in all cases
are attractive, due to the strong attraction provided by $\sigma$
exchange. In the $I$=0 $C$=+ case, the spin factor
$<\sigma_q\cdot\sigma_{\bar{q}}>$ in eqs. (\ref{eff:pseudoscalar},
\ref{eff:vector}) is 1, and the attractive interaction potential, as
shown in Fig. \ref{SP:SV}, is especially strong, because $\pi$  and
$\sigma'$ exchanges also provide strong attraction.

Not only in this $I$=0 $C$=+ case but also in the case of $I$=0
$C$=-- can we find the bound states after solving the
Schr\"{o}dinger equation. We list the obtained binding energies in
Table \ref{SPSV} and one may find the binding energies, for the
systems of $D\bar{D}^*_0$, $D\bar{D}^*_2$ and $D^*\bar{D}_1$, are
almost the same, and the small differences are due to the small
differences among their reduced masses.

\begin{table}[htbp]\renewcommand\arraystretch{1.5}
\caption{Binding energies (MeV) of the bound states of S-P D meson
systems.  I, II and III refer to the chiral SU(3) quark model, the
extended chiral SU(3) quark model neglecting and considering tensor
coupling of vector field, respectively.}\label{SPSV} \centering
\begin{tabular*}{8cm}{@{\extracolsep{\fill}}ccccc}
\hline
&$I^G(J^{PC})$& I & II & III  \\\hline
$D\bar{D}_1$&$0^+(1^{-+})$&0.4&32.4&22.6\\
&$0^-(1^{--})$&0.3&21.0&12.0\\
 \hline
$D\bar{D}^*_0$&$0^+(0^{-+})$&1.7&29.3&19.6\\
&$0^-(0^{--})$&0.4&23.2&15\\
$D\bar{D}^*_2$&$0^+(2^{-+})$&1.8&29.7&20.0\\
&$0^-(2^{--})$&0.5&23.6&15.4\\
$D^*\bar{D}_1$&$0^+(J^{-+})$&2.2&31.2&21.2\\
&$0^-(J^{--})$&0.7&24.7&16.3\\
 \hline
 $D^*\bar{D}^*_0$&$0^+(1^{-+})$(S=2)&21.8&30.7&30.3\\

&$0^-(1^{--})$(S=2)&5.9&12.6&12.1\\

$D^*\bar{D}^*_2$&$0^+(J^{-+})$(S=2)&22.3&31.3&30.9\\

&$0^-(J^{--})$(S=2)&6.1&12.9&12.4\\
\hline
\end{tabular*}
\end{table}

We find the $0^+(0^{-+})$ $D\bar{D}^*_0$ molecule has a mass of
4154-4181 MeV, and it could be a plausible explanation for the
resonance of $X(4160)$. The mass of $0^-(0^{--})$ $D\bar{D}^*_0$
molecule, in our calculation, is about 4160-4183 MeV. Moreover, we
get the $0^+(2^{-+})$ $D\bar{D}^*_2$ molecule with the mass of
4297-4325 MeV, and it could be an explanation for $X(4350)$. The
$0^-(2^{--})$ $D\bar{D}^*_2$ molecule has a mass of 4303-4326 MeV,
and the $0^+(J^{-+})$ $D^*\bar{D}_1$ or $0^-(J^{--})$ $D^*\bar{D}_1$
molecules has a mass of 4400-4429 MeV or 4406-4430 MeV,
respectively. However, no $I$=1 $D^*\bar{D}_1$ molecule could be
found in our model, because $\pi$ and $\sigma^{'}$ exchanges are both
repulsive. Therefore, our result doesn't favor the $D^*\bar{D}_1$
molecular explanation of the isovector $X(4430)^{\pm}$. This result
is consistent with the conclusion of Liu \cite{liu0}, and contrary
to the results of Close \cite{close} and Ding \cite{ding1}.

\begin{figure}[htb]
\centering
\includegraphics[scale=0.3]{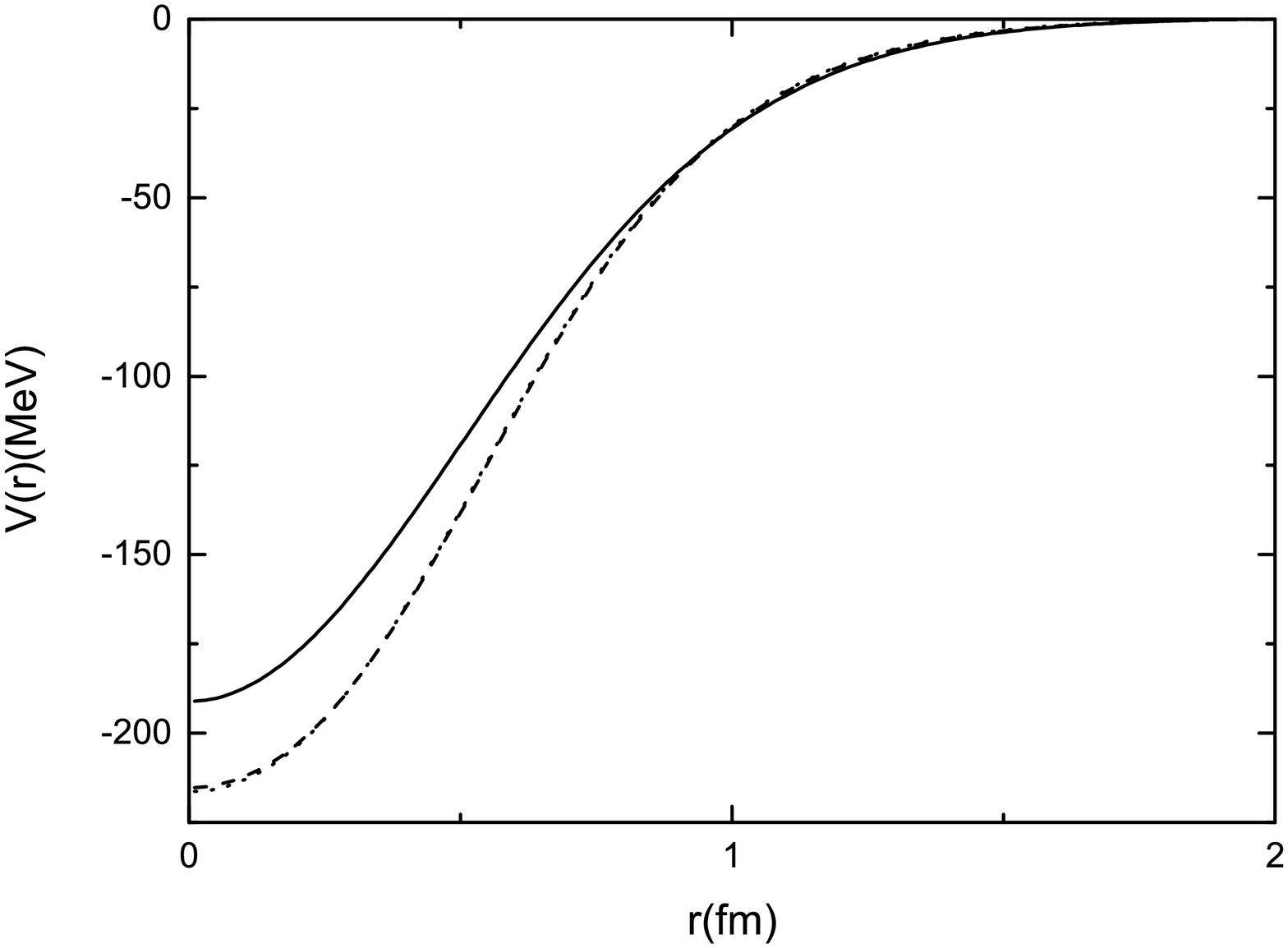}
\caption{The total interaction potential of $I$=0 $C$=+ $S$=2
$D^*\bar{D}^*_0$ and $D^*\bar{D}^*_2$ systems.   The solid, dashed
and dotted lines represent the same meaning as in Fig. \ref{SP:SS}.
\label{SP:VV} }
\end{figure}

For the two different vector meson systems, such as $D^*\bar{D}^*_0$
and $D^*\bar{D}^*_2$, their interaction potentials  are
simultaneously associated with the isospin $I$, the $C$-parity $C$
and the total spin $S$. In the $I$=0, $C$=+, $S$=2 case, the spin
factor $<\sigma_q\cdot\sigma_{\bar{q}}>$, in eqs.
(\ref{eff:pseudoscalar}, \ref{eff:vector}), is 2, which is different
from that of $I$=0 $C$=+ of the $D\bar{D}^*_0$, $D\bar{D}^*_2$ and
$D^*\bar{D}_1$ systems. We find that the $D^*\bar{D}^*_0$ and
$D^*\bar{D}^*_2$ systems share the same interaction potential shown
in Fig. \ref{SP:VV}, and correspondingly we get the bound states
with the binding energies listed in the  Table \ref{SPSV}. In this
case the effects of the vector meson exchanges are small in the extended chiral
SU(3) quark model. In the case of $I$=0, $C$=--, $S$=2, the bound
states also exist.

In our calculation, the mass of $0^+(1^{-+})$(S=2) $D^*\bar{D}^*_0$
molecule is of 4297-4306MeV and we find the $0^-(1^{--})$(S=2)
$D^*\bar{D}^*_0$ molecule has a mass of 4315-4322MeV, and it might
be explained as $X(4360)$. No bound states could be found in the
$I$=1 cases in our approach.  Our result agrees with Ding
\cite{ding} that $X(4250)^{\pm}$ might not be explained as a
isovector $D^*\bar{D}^*_0$ molecule. Our results also disfavor the
$D^*\bar{D}^*_0$ molecular explanation of $X(4260)$ \cite{nielsen,
nielsen1,nielsen2}. Since we ignore the spin-orbital coupling
interaction in our calculation, the $0^+(J^{-+})$(S=2) $D^*\bar{D}
^*_2$  molecule with mass of 4441-4452MeV and the $0^-(J^{--})$(S=2)
$D^*\bar{D} ^*_2$ molecule with mass of 4459-4466MeV are degenerate
states for the total angular momentum $J$.

\subsubsection{P-P: $D_1\bar{D}_1$, ($D_1\bar{D}^*_0$,
$D_1\bar{D}^*_2$) and  ($D^*_0\bar{D}^*_0$, $D^*_2\bar{D}^*_2$,
$D^*_0\bar{D}^*_2$)}

\begin{figure}[htb]
\centering
\includegraphics[scale=0.3]{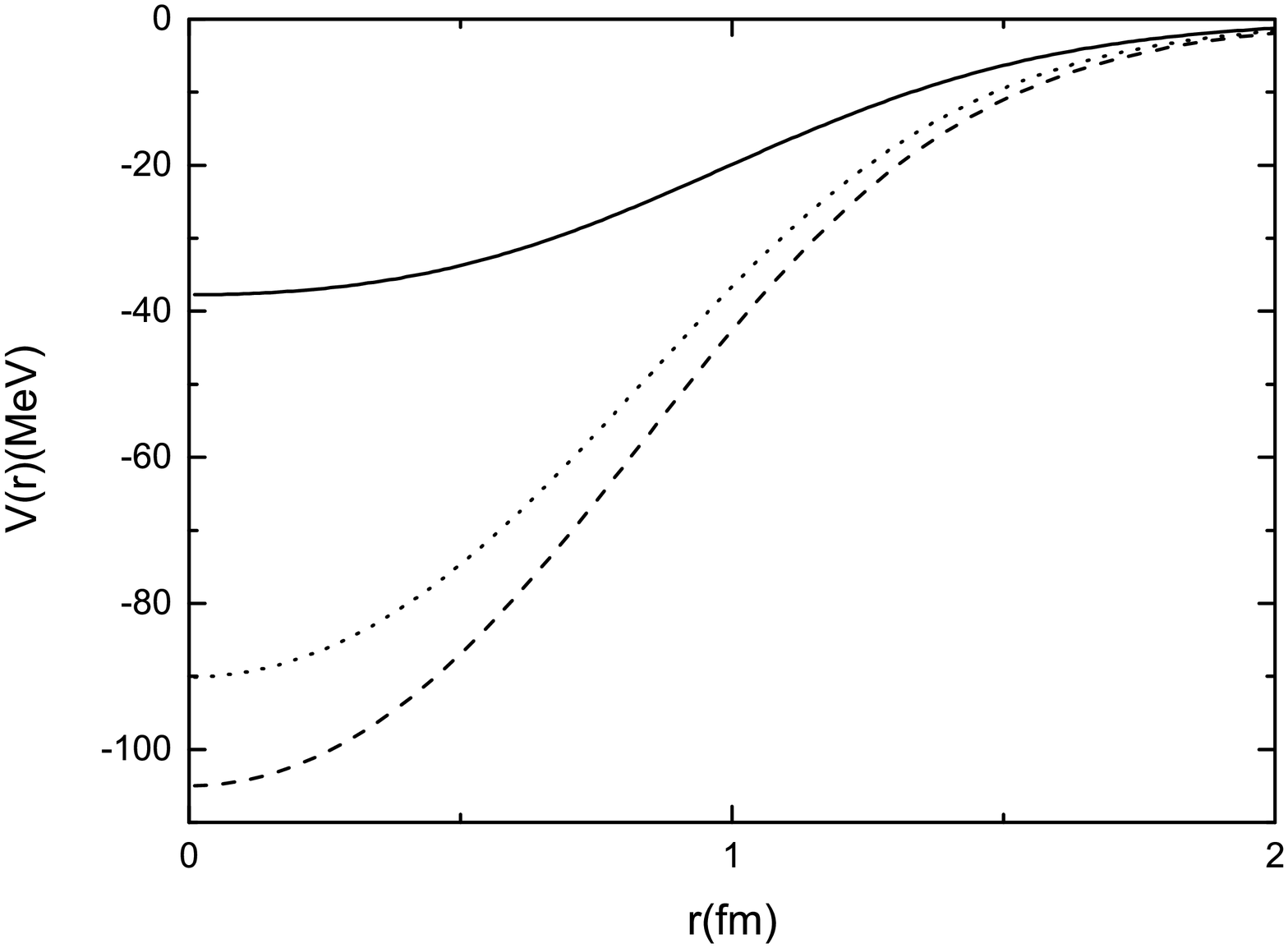}
\caption{The total interaction potential of  $I$=0
$D_1\bar{D}_1$ system. The solid, dashed
and dotted lines represent the same meaning as in Fig. \ref{SP:SS}. \label{PP:SS} }
\end{figure}

For the system of $D_1\bar{D}_1$, its interaction is only associated
with isospin $I$. It could form a $0^+(J^{++})$ weakly bound state
with binding energy listed in Table \ref{PPSV}. This bound state has
a mass of 4824-4842 MeV in our approach. The interaction potential
is shown in Fig. \ref{PP:SS}.

\begin{figure}[htb]
\centering
\includegraphics[scale=0.3]{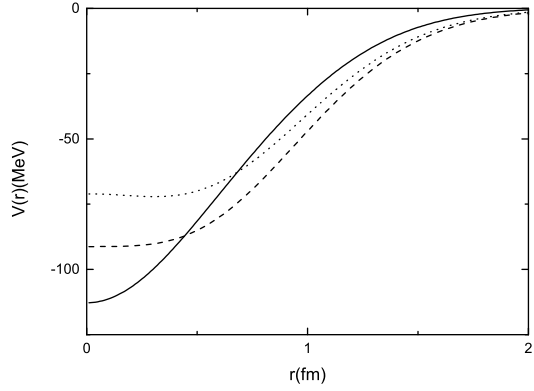}
\caption{The total interaction potential of $I$=0 and $C$=+
$D_1\bar{D}^*_0$ and $D_1\bar{D}^*_2$ systems and $I$=0
$S$=2 $D^*_0\bar{D}^*_0$,  $D^*_2\bar{D}^*_2$ and $D^*_0\bar{D}^*_2$ systems.  The solid, dashed
and dotted lines represent the same meaning as in Fig. \ref{SP:SS}.
\label{PP:SV} }
\end{figure}

For the $D_1\bar{D}^*_0$ and $D_1\bar{D}^*_2$ systems with the same
quantum numbers, they share the same interaction potential. Also in
the $I$=0 $C$=+ case the interaction potential becomes strong enough
to form bound states. This interaction potential is shown in Fig.
\ref{PP:SV}.
We see that the interaction potential in the extended chiral SU(3)
quark model is weaker than that in the chiral SU(3) quark model,
because the vector meson exchanges, in this case, provide repulsive
interaction, however the range of the total potential is longer than
that of the chiral SU(3) quark model. As a result, the vector meson
exchanges don't obviously change the binding energies as listed in
Table \ref{PPSV}. In our approach, the $0^+(1^{++})$
$D_1\bar{D}^*_0$ and
 $0^+(J^{++})$ $D_1\bar{D}^*_2$  molecules have the masses of
 4719-4727 MeV and 4863-4870 MeV, respectively.

\begin{table}[htbp]\renewcommand\arraystretch{1.5}
\caption{Binding energies (MeV) of the bound states of P-P D meson
systems.  I, II and III refer to the same meaning as in Table
\ref{SPSV}.}\label{PPSV} \centering
\begin{tabular*}{8cm}{@{\extracolsep{\fill}}ccccc}
\hline
&$I^G(J^{PC})$& I & II & III  \\\hline
$D_1\bar{D}_1$&$0^+(J^{++})$&0.4&18.2&12.0\\
\hline
$D_1\bar{D}^*_0$&$0^+(1^{++})$&12.2&19.5&12.4\\

$D_1\bar{D}^*_2$&$0^+(J^{++})$&12.5&19.8&12.4\\

\hline
$D^*_0\bar{D}^*_0$&$0^+(0^{++})$(S=2)&12.1&19.3&12.4\\
&$0^-(0^{+-})$(S=1)&--&16.7&11.4\\
&$0^+(0^{++})$(S=0)&--&15.6&11.3\\

$D^*_2\bar{D}^*_2$&$0^+(J^{++})$(S=2)&12.7&20.0&12.9\\
&$0^-(J^{+-})$(S=1)&--&17.5&12.1\\
&$0^+(J^{++})$(S=0)&--&16.3&11.9\\

  $D^*_0\bar{D}^*_2$&$0^+(2^{++})$(S=2)&12.4&19.7&12.6\\
 & $0^+(2^{++})$(S=1)&--&17.1&11.7\\
&  $0^+(2^{++})$(S=0)&--&16.0&11.6\\

\hline
\end{tabular*}
\end{table}

For the two vector systems of $D^*_0\bar{D}^*_0$,
$D^*_2\bar{D}^*_2$ and $D^*_0\bar{D}^*_2$, they share the same interaction corresponding to
the same isospin $I$ and total spin $S$, and the possible bound
states are also listed in Table \ref{PPSV}.
In the case of $I$=0
$S$=2 $D^*_0\bar{D}^*_0$, $D^*_2\bar{D}^*_2$ and $D^*_0\bar{D}^*_2$
systems, the spin factor $<\sigma_q\cdot\sigma_{\bar{q}}>$ in eqs.
(\ref{eff:pseudoscalar}, \ref{eff:vector}) is 1, the same as that of
the $I$=0 $C$=+ $D_1\bar{D}^*_0$ and $D_1\bar{D}^*_2$  systems. Here
the interaction potential is shown in Fig. \ref{PP:SV}, which is
strong enough to form the bound states in our models. Our results
tell the possible $0^+(0^{++})$ $D^*_0\bar{D}^*_0$ molecule with the
mass of 4617-4624 MeV, the $0^+(J^{++})$ $D^*_2\bar{D}^*_2$ molecule
with the mass of 4904-4911 MeV and the possible $0^+(0^{++})$
$D^*_0\bar{D}^*_2$ molecule has a mass of 4760-4768 MeV.

For the systems of $I$=0 $S$=0,1  $D^*_0\bar{D}^*_0$,
$D^*_2\bar{D}^*_2$ and $D^*_0\bar{D}^*_2$, we get the bound states
in our extended chiral SU(3) quark model, but no bound states in our
chiral SU(3) quark model. This is because, in our extended chiral
SU(3) quark model, vector meson exchanges provide additional strong
attractive interaction. Even though we can't currently draw a
definite conclusion whether or not the molecular bound states could
exist, those states provide a test for our two models.

As a short summary, we have studied twelve sets of the heavy meson
interactions associated with P-wave D mesons. They are,
S-P: $D\bar{D}_1$, ($D\bar{D}^*_0$, $D\bar{D}^*_2$, $D^*\bar{D}_1$) and
($D^*\bar{D}^*_0$, $D^*\bar{D}^*_2$);
and P-P: $D_1\bar{D}_1$, ($D_1\bar{D}^*_0$,
$D_1\bar{D}^*_2$) and  ($D^*_0\bar{D}^*_0$, $D^*_2\bar{D}^*_2$,
$D^*_0\bar{D}^*_2$). We've found eighteen
possible bound states, and they are all isoscalar. Only the four
possible molecules of $0^+(0^{-+})$ $D\bar{D}^*_0$, $0^-(1^{--})$
$D^*\bar{D}^*_0$, $0^-(1^{--})$ $D\bar{D}_1$, and $0^+(2^{-+})$
$D\bar{D}^*_2$ could be explained as $X(4160)$, $X(4360)$,
$X(4260)$, $X(4350)$, respectively. Other bound states don't match
any observed XYZ states.

\subsection{Two B masons}

Considering the resemblance between B mesons and D mesons, the
corresponding binding systems should have the similar properties,
while the P-wave B meson bound states should have larger binding
energies because of their larger reduced masses.

\subsubsection{S-P: $B\bar{B}_1$, ($B\bar{B}^*_2$, $B_1\bar{B}^*$) and
$B^*\bar{B}^*_2$}
\begin{figure}[htb]
\centering
\includegraphics[scale=0.3]{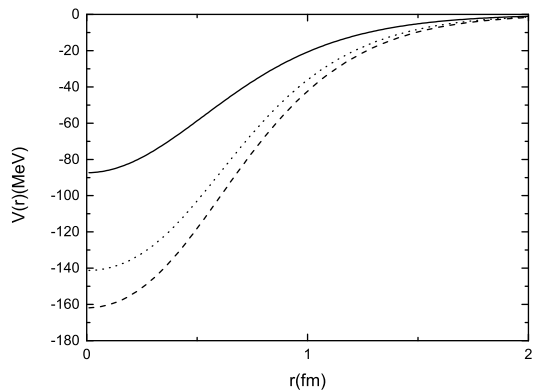}
\caption{The total interaction potential of  $I$=0 $C$=+
$B\bar{B}^*_2$ and $B_1\bar{B}^*$ systems.  The solid, dashed
and dotted lines represent the same meaning as in Fig. \ref{SP:SS}.
\label{SP:SVB} }
\end{figure}

 As for the $B\bar{B}_1$ system, similar to the $D\bar{D}_1$ system, it becomes
bound in the  $I=0$ $C=\pm$ cases and the obtained binding energies
are listed in the Table \ref{SPSVB}.
In the $I=0$ $C=+$ case, the $0^+(1^{-+})$ $B\bar{B}_1$ molecule has a mass of 10932-10980 MeV,
and in the $I=0$ $C=-$ case, the $0^-(1^{--})$ $B\bar{B}_1$ molecule has a mass of 10955-10993 MeV,

For $B\bar{B}^*_2$ and $B_1\bar{B}^*$, the systems of one S-wave B
meson and one P-wave B meson, they share the same interaction
potential corresponding to the same isospin and $C$-parity. In the
$I$=0 case the interaction potential is strong enough to bind
$B\bar{B}^*_2$ and $B_1\bar{B}^*$. For example, in the $I$=0 and
$C$=+ case, the interaction potential is shown in Fig. \ref{SP:SVB},
and the obtained binding energies are listed in Table \ref{SPSVB}.
\begin{table}[htbp]\renewcommand\arraystretch{1.5}
\caption{Binding energies (MeV) of the bound states of S-P B meson
systems.  I, II and III refer to the same meaning as that in Table
\ref{SPSV}.}\label{SPSVB} \centering
\begin{tabular*}{8cm}{@{\extracolsep{\fill}}ccccc}
\hline
&$I^G(J^{PC})$& I & II & III  \\
 \hline
 $B\bar{B}_1$&$0^+(1^{-+})$&20.0&69.0&56.1\\
 &$0^-(1^{--})$&7.7&46.1&34.5\\
\hline
$B\bar{B}^*_2$&$0^+(2^{-+})$&17.8&64.6&51.2\\
&$0^-(2^{--})$&9.2&50.1&38.6\\
$B_1\bar{B}^*$&$0^+(J^{-+})$&17.8&64.7&51.3\\
&$0^-(J^{--})$&9.2&50.2&38.6\\
 \hline
$B^*\bar{B}^*_2$& $0^+(J^{-+})$(S=2)&66.8&97.8&91.7\\
& $0^+(J^{-+})$(S=1)&2.2&9.4&7.7\\

&$0^-(J^{--})$(S=2)&24.9&40.8&40.1\\
\hline

 \end{tabular*}
\end{table}
We find that the $0^+(2^{-+})$ $B\bar{B}^*_2$ and $0^-(2^{--})$
$B\bar{B}^*_2$ molecules have the mass of 10962-11009 MeV and
10977-11018 MeV, respectively. In our calculation the $0^+(J^{-+})$
$B_1\bar{B}^*$ molecule, without a definite total angular momentum
$J$, has a mass of 10970-11017 MeV, and the mass of  $0^-(J^{--})$
$B_1\bar{B}^*$ molecule is about 10985-11026 MeV.

\begin{figure}[htb]\renewcommand\arraystretch{1.5}
\centering
\includegraphics[scale=0.3]{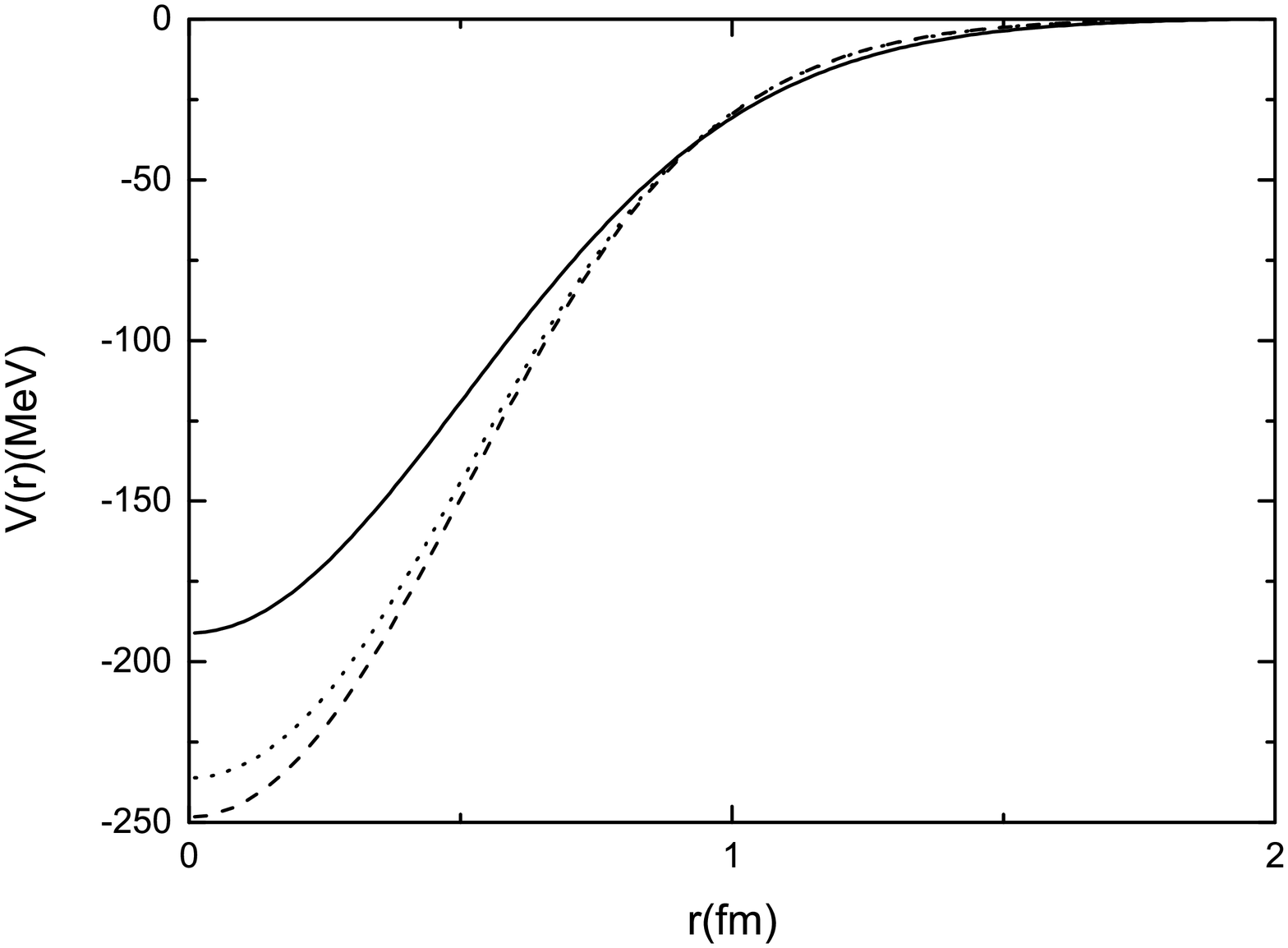}
\caption{The total interaction potential of $I$=0 $C$=+ $S$=2
$B^*\bar{B}^*_2$ system.  The solid, dashed
and dotted lines represent the same meaning as in Fig. \ref{SP:SS}.\label{PV:SVB2} }
\end{figure}

For the $B^*\bar{B}^*_2$ system, the total interaction potential is
associated with isospin $I$, $C$-parity $C$=+ and total spin $S$.
The possible bound states are listed in the Table \ref{SPSVB}. The
interaction potential between $B^*\bar{B}^*_2$ and in the case of
$I$=0 $C$=+ $S$=2 is shown in Fig. \ref{PV:SVB2}.
This interaction potential is strong enough to bind $B^*\bar{B}^*_2$
system, and the $0^+(J^{-+})$(S=2) $B^*\bar{B}^*_2$ molecule has a
mass of 10974-11005 MeV in our calculation. In the case of $I$=0
$C$=+ $S$=1, the mass of $0^+(J^{-+})$(S=1) $B^*\bar{B}^*_2$
molecule is about 11063-11070 MeV, while in the case of $I$=0 $C$=-
$S$=2, the $0^-(J^{--})$(S=2) $B^*\bar{B}^*_2$ molecule has a mass
of 11061-11047 MeV.

\subsubsection{P-P: $B_1\bar{B}_1$, $B_1\bar{B}^*_2$ and
$B^*_2\bar{B}^*_2$}
For the two pseudoscalar meson system of
$B_1\bar{B}_1$, the total spin $S$ is 0, like $D_1\bar{D}_1$, the
total interaction potential is only associated with the isospin $I$.
Like the $D_1\bar{D}_1$ system, $B_1\bar{B}_1$ forms a bound state
in the $I$=0 case in our models as shown in Table \ref{PPSVB}. Our
$0^+(J^{++})$ $B_1\bar{B}_1$ molecule has a mass of 11408-11436 MeV.

\begin{figure}[htb]
\centering
\includegraphics[scale=0.3]{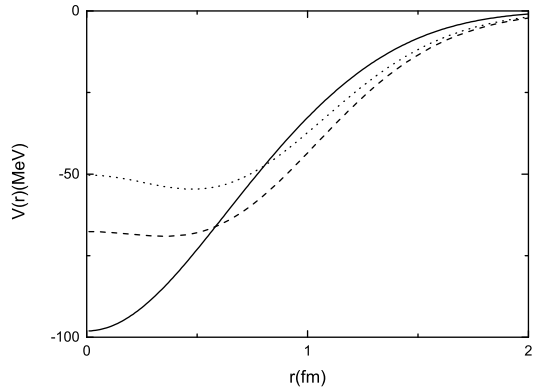}
\caption{The total interaction potential of $I$=0 and $C$=+
$B_1\bar{B}^*_2$ system and $I$=0 $S$=2 $D^*_2 \bar{D}^*_2$ system.
The solid, dashed
and dotted lines represent the same meaning as in Fig. \ref{SP:SS}.
\label{PP:SVB} }
\end{figure}

For the $B_1\bar{B}^*_2$, like the $D_1\bar{D}^*_2$ system, the
interaction potential shown in Fig. \ref{PP:SVB}, in the case of
$I$=0 and $C$=+, is strong to bind the $B_1\bar{B}^*_2$.
The obtained binding energies in this case are listed in the Table
\ref{PPSVB}.
\begin{table}[htbp]\renewcommand\arraystretch{1.5}
\caption{Binding energies (MeV) of the bound states of P-P B meson systems.
 I, II and III
refer to the same meaning as in Table \ref{SPSV}.}\label{PPSVB}
\centering
\begin{tabular*}{8cm}{@{\extracolsep{\fill}}ccccc}
\hline
&$I^G(J^{PC})$& I & II & III  \\
 \hline
  $B_1\bar{B}_1$&$0^+(J^{++})$&7.1&39.5&30.5\\
 \hline
$B_1\bar{B}^*_2$&$0^+(J^{++})$&31.1&33.2&24.3\\
&$0^-(J^{+-})$&--&20.2&14.2\\
\hline
$B^*_2\bar{B}^*_2$&$0^+(J^{++})$(S=2)&31.2&32.5&23.7\\
&$0^-(J^{+-})$(S=1)&--&36.8&30.4\\
&$0^+(J^{++})$(S=0)&--&40.1&35.7\\
\hline

 \end{tabular*}
\end{table}
We find that the  $0^+(J^{++})$ $B_1\bar{B}^*_2$ molecule has a mass
of 11435-11444 MeV.

As for the system of $B^*_2\bar{B}^*_2$, the interaction potential
is associated with isospin $I$ and the total spin $S$. Similar to
the $D^*_2\bar{D}^*_2$ system, we find a bound state in the $I$=0
$S$=2 case where the interaction potential is just the same as the
$I$=0 $C$=+ $B_1\bar{B}^*_2$ system shown in Fig. \ref{PP:SVB}. The
binding energies are also listed in the Table \ref{PPSVB}. In our
approach, the $0^+(J^{++})$(S=2) molecule has a mass of 11454-11465
MeV.

To summarize this subsection, we have studied seven sets of heavy
meson interactions associated with P-wave B mesons. They are S-P:
$B\bar{B}^*_2$, ($B^*\bar{B}_1$, $B\bar{B}_1$), and $B^*\bar{B}^*_2$
and P-P: $B_1\bar{B}^*_2$, $B^*_2\bar{B}^*_2$, and $B_1\bar{B}_1$.
We've found twelve possible bound states and they are all isoscalar.
Unfortunately, none of these bound states could match any newly
observed XYZ states.

\section{Summary}\label{sec:sum}

In this work, we have employed our SU(3) chiral quark model to
systematically studied the heavy meson interactions associated with
the P-wave. The parameters of our models have already fixed in our
previous calculations from the spectra of baryon ground states, the
nucleon-nucleon phase shifts and  the deuteron binding energy. We
have gotten some possible molecular states. For the systems of
P-wave B-mesons, unfortunately, those molecules don't match to any
of the newly observed XYZ hadronic states. For P-wave D meson
interactions, we find that the $X(4350)$ could be explained as the
$0^+(2^{-+})$ $D\bar{D}^*_2$ molecule, the $X(4360)$ could be
explained as the $0^-(1^{--})$ $D^*\bar{D}^*_0$ molecule,  and the
$X(4160)$ could be explained as the $0^+(1^{-+})$ $D\bar{D}^*_0$
molecule. However, the $X(4430)^{\pm}$ could not be explained as a
isovector $D_1D^*$ charged molecule. Moreover, neither
$X(4250)^{\pm}$ nor $X(4260)$ could be explained as the
$D^*\bar{D}^*_0$ molecule, but $X(4260)$ might be explained as a
$0^- (1^{--})$ $D\bar{D}_1$ molecule.

Finally, we find that we can get bound states in our extended chiral
SU(3) quark model in some cases but not in the chiral SU(3) quark
model. This is because the vector meson exchanges in the extended
chiral SU(3) quark model provide additional attraction. Future
experimental results about these special cases would give a
discriminates of our two models.

\begin{acknowledgments}

This project is supported, in part, by National Natural Science
Foundation of China (Grant Nos. 11105158, 11035006,10975146,
11261130), by Ministry of Science and Technology of China (Grant No.
2009CB825200). This work is also supported by the DFG and NSFC
through funds provides to the Sino-German CRC 110 "Symmetries and
the emergence of structure in QCD". M.T. Li thanks Prof.
Sch\"{o}berl for his kindly help in solving Schr\"{o}dinger Equation
\cite{schoberl1985,schoberl1999}.

\end{acknowledgments}


\begin{thebibliography}{99}
\bibitem{3872-Belle}Belle Collaboration, S.K. Choi et al., Phys. Rev. Lett. {\bf 91}, 262001 (2003).
\bibitem{3872-CDF}CDF Collaboration, D. Acosta et al., Phys. Rev. Lett. {\bf 93}, 072001 (2004).
\bibitem{3872-D0}D0 Collaboration, V.M. Abazov et al., Phys. Rev. Lett. {\bf 93}, 162002 (2003).
\bibitem{3872-Babar}Babar Collaboration, B. Aubert et al., Phys. Rev. D {\bf 71}, 071103
(2005).

\bibitem{x3940-Belle}Belle Collaboration, K. Abe et al., Phys. Rev. Lett. {\bf 98}, 082001 (2007).
\bibitem{y3940-Belle}Belle Collaboration, S.K. Choi et al., Phys. Rev. Lett. {\bf 94}, 182002 (2005).
\bibitem{y3940-Babar}Babar Collaboration, B. Aubert et al., Phys. Rev. Lett. {\bf 101}, 082001 (2008).

\bibitem{4050-Babar}Babar Collaboration, J. P. Lees et al., Phys. Rev. D {\bf 85}, 052003 (2012).
\bibitem{4050-Belle}Belle Collaboration, R. Mizuk et al., Phys. Rev. D {\bf 78}, 072004 (2008).

\bibitem{4140-Belle}Belle Collaboration, C.P. Shen et al., Phys. Rev. Lett. {\bf 104}, 112004 (2010).
\bibitem{4140-CDF}CDF Collaboration, T. Aaltonen et al., Phys. Rev. Lett. {\bf 102}, 242002 (2009).
\bibitem{X4160-Belle}Belle Collaboration, P. Pakhlov et al., Phys. Rev. Lett. {\bf 100}, 202001 (2008).

\bibitem{4260-Babar}Babar Collaboration, B. Aubert et al., Phys. Rev. Lett. {\bf 95}, 142001 (2005).
\bibitem{4260-Cleo}CLEO Collaboration, T.E. Coan et al., Phys. Rev. Lett. {\bf 96}, 162003 (2006).
\bibitem{4260-Belle}Belle Collaboration, C.Z. Yuan et al., Phys. Rev. Lett. {\bf 99}, 182004 (2007).

\bibitem{4360-Belle}Belle Collaboration, X. L. Wang et al., Phys. Rev. Lett. {\bf 99}, 142002 (2007).
\bibitem{4360-Babar}Babar Collaboration, B. Aubert et al., Phys. Rev. D {\bf 98}, 212001 (2007).

\bibitem{4430-Belle}Belle Collaboration, S.-K. Choi et al., Phys. Rev. Lett. {\bf 100}, 142001 (2008).
\bibitem{4430-Belle-again}Belle Collaboration, R. Mizuk et al., Phys. Rev. D {\bf 80}, 031104 (2009).
\bibitem{4430-Babar}Babar Collaboration, B. Aubert et al., Phys. Rev. D {\bf 79}, 112001 (2009).

\bibitem{Zb-Belle}Belle Collaboration, A. Bondar et al., Phys. Rev. Lett. {\bf 108}, 122001 (2012).
\bibitem{li1}M.T. Li, W.L. Wang, Y.B. Dong, Z.Y. Zhang,  J. Phys. G \textbf{40}, 015003 (2013).
\bibitem{li2}M.T. Li, W.L. Wang, Y.B. Dong, Z.Y. Zhang, Int. J. Mod. Phys. A \textbf{27} 1250161 (2012), arxiv:1206.0523[hep-ph].

\bibitem{close}F. Close, C. Downum, C. E. Thomas, Phys. Rev. D \textbf{81}, 074033 (2010).

\bibitem{nielsen}R.M. Albuquerque, M. Nielsen,  Nucl. Phys. A \textbf{857}, 48-49 (2011).
\bibitem{nielsen1}S.H. Lee, K. Moritaa, , M. Nielsen, Nucl. Phys. A \textbf{815}, 29-39 (2009).
\bibitem{nielsen2}M. Nielsen, F.S. Navarraa, S.H. Lee, Phys. Rep. \textbf{497}, 41-83 (2010).
\bibitem{ding}G.J. Ding, Phys. Rev. D \textbf{79}, 014001 (2009).
\bibitem{ding1}G.J. Ding, W. Huang, J.F. Liu, M.L. Yan, Phys. Rev. D \textbf{79}, 034026 (2009).
\bibitem{liu0}Y.R. Liu, Z.Y. Zhang, arxiv: 0908.1734[hep-ph].
\bibitem{zhang1}Z.Y. Zhang, A. Faessler, U. Straub, L.Ya. Glozman, Nucl. Phys. A \textbf{578}, 573 (1994).
\bibitem{zhang2}Z.Y. Zhang, Y.W. Yu, P.N. Shen, L.R. Dai, A. Faessler, and U. Straub, Nucl. Phys. A  \textbf{625}, 59 (1997).
\bibitem{dai}L.R. Dai, Z.Y. Zhang, Y.W. Yu, P. Wang, Nucl. Phys. A \textbf{727}, 321 (2003).
\bibitem{fhuang04kn}F. Huang, Z.Y. Zhang, and Y.W. Yu, Phys. Rev. C {\bf 70}, 044004 (2004).
\bibitem{fhuang04nkdk}F. Huang and Z.Y. Zhang, Phys. Rev. C {\bf 70}, 064004 (2004).
\bibitem{liu1}Y.R. Liu and Z.Y. Zhang, Phys. Rev. C \textbf{80}, 015208 (2009).
\bibitem{liu2}Y.R. Liu and Z.Y. Zhang, Phys. Rev. C \textbf{79}, 035206 (2009)
\bibitem{wangwlsigc}W.L. Wang, F. Huang, Z.Y. Zhang, and B.S. Zou, Phys. Rev. C  \textbf{84}, 015203 (2011).
\bibitem{li}M.T. Li, Y.B. Dong, Z.Y. Zhang, Chin.Phys. C \textbf{35} 622-628 (2011), arxiv: 1010.2283[hep-ph].

\bibitem{wang2007}W.L. Wang, F. Huang, Z.Y. Zhang, Y.W. Yu and F. Liu, Eur. Phys. J. A \textbf{32}, 293-297 (2007).
\bibitem{wang2008}W.L. Wang, F. Huang, Z.Y. Zhang and F. Liu, J. Phys. G \textbf{35}, 085003 (2008).
\bibitem{wang2010}W.L. Wang, F. Huang, Z.Y. Zhang and F. Liu, Mod. Phys. Lett. A \textbf{25}, 1325-1332 (2010).
\bibitem{wang2011}W.L. Wang and Z.Y. Zhang, Phys. Rev. C \textbf{84}, 054006 (2011).

\bibitem{PDG2010}K. Nakamura {\it et al.} (Particle Data Group), J. Phys. G 37, 075021 (2010).

\bibitem{zhanghx1}H.X. Zhang ,W.L. Wang, Y.B. Dai, and Z.Y. Zhang, Commun. Theor. Phys. \textbf{49}, 414 (2008).
\bibitem{zhanghx2}H.X. Zhang, M. Zhang, and Z.Y. Zhang, Chin. Phys. Lett. \textbf{24}, 2533 (2007); M. Zhang, H.X. Zhang, and Z.Y. Zhang, Commun. Theor. Phys. \textbf{50}, 437 (2008).
\bibitem{liux}X. Liu, Z. G. Luo, Y.R. Liu and S.L. Zhu, Eur. Phys. J. C 61, 411 (2009).
\bibitem{he}Z.F. Sun, J. He, X. Liu, Z.G. Luo and S.L. Zhu, Phys. Rev. D \textbf{84}, 054002 (2011).


\bibitem{schoberl1985}P. Falkensteiner, H. Grosse, F. Sch\"{o}berl and P. Hertel, Comput. Phys. Commun. \textbf{34}, 287-293 (1985).
\bibitem{schoberl1999}W. Lucha and F. F. Sch\"{o}berl, Int. J. Mod. Phys. C \textbf{10}, 607-619 (1999).

\end{thebibliography}
\end{document}